\documentclass[byrevtex,superscriptaddress,twocolumn,prl]{revtex4}%
\usepackage{amsfonts}
\usepackage{amsmath}
\usepackage{amssymb}
\usepackage{graphicx}%
\begin{document}
\preprint{ }
\title{Elimination of the Diffraction of Arbitrary Images Imprinted on Slow Light}
\author{O. Firstenberg}
\affiliation{Department of Physics, Technion-Israel Institute of Technology, Haifa 32000, Israel}
\author{M. Shuker}
\affiliation{Department of Physics, Technion-Israel Institute of Technology, Haifa 32000, Israel}
\author{N. Davidson}
\affiliation{Department of Physics of Complex Systems, Weizmann Institute of Science,
Rehovot 76100, Israel}
\author{A. Ron}
\affiliation{Department of Physics, Technion-Israel Institute of Technology, Haifa 32000, Israel}


\begin{abstract}
We present a scheme for eliminating the optical diffraction of slow-light in a
thermal atomic medium of electromagnetically induced transparency.
Nondiffraction is achieved for an arbitrary paraxial image by manipulating the
susceptibility in momentum space, in contrast to the common approach, which
employs guidance of specific modes by manipulating the susceptibility in real
space. For negative two-photon detuning, the moving atoms drag the transverse
momentum components unequally, resulting in a Doppler trapping of light by
atoms in two dimensions.

\end{abstract}
\maketitle

Every classical wave field is subjected to diffraction throughout its
propagation. Nondiffracting beams, \textit{i.e.}, optical modes that maintain
their intensity distribution in the transverse planes normal to the
propagation direction, exist only within the particular class of Bessel beams
\cite{DurninJOSA1987}. There is no solution, in any field in optics, to
suppress diffraction for an arbitrary image and for any distance along the
propagation direction.

In recent years, the process of electromagnetically induced transparency (EIT)
has been employed to reduce or eliminate the diffraction spreading of beamlike
fields \nocite{MoseleyPRL1995}\nocite{MitsunagaPRA2000}%
\nocite{WilsonGordonPRA1998}\nocite{AgarwalPRA2000,TruscottPRL1999}%
\nocite{LukinPRL2005_strong_confinement}\nocite{chengPRA2005}%
\nocite{ChengOL2007}\nocite{VengalattorePRL2005,TarhanOL2007}%
\nocite{HongPRL2003,FriedlerOL2005} [2-13] by manipulating the susceptibility
in \emph{real space} and inducing a gradient of the index of refraction
\cite{GomezReinoAO1982}. Similarly to waveguiding, special modes, such as the
Laguerre-Gauss modes, propagate in the induced waveguides without diffraction
or, equivalently, arbitrary images can be revived after a certain self-imaging
distance. In this paper, we suggest a method to achieve light propagation
without diffraction for any arbitrary paraxial image, with both the intensity
and phase information of the image completely maintained. We utilize Dicke
narrowing in a vapor EIT medium \cite{FirstenbergPRA2007} to obtain a
susceptibility that is quadratic in the \emph{transverse momentum space} and
by that eliminate the effect of diffraction. A unique manifestation of
nondiffraction is the ability to suspend the expansion of a beam regardless of
its position. Other applications may include high-resolution imaging, slowing
and storage of images \cite{ShukerImaging2007,HowellPRL2008}, and nonlinear
optics \cite{LukinPRL2005_strong_confinement}.

In EIT, a beamlike probe field traverses the medium with a reduced group
velocity, in the presence of a second pump field. Spatial manipulation of the
probe's susceptibility may be achieved either by applying a suitable
nonuniform pump beam or by employing inhomogeneity of the atomic medium. The
former technique, known as \emph{electromagnetically induced focusing}, was
observed in a vapor medium \cite{MoseleyPRL1995} and later with cold atoms
\cite{MitsunagaPRA2000}. Exact cancellation of diffraction by induced focusing
was studied extensively as induced solitons \cite{WilsonGordonPRA1998},
induced waveguides \cite{AgarwalPRA2000,TruscottPRL1999}, and transverse
confinement \cite{LukinPRL2005_strong_confinement,chengPRA2005}, but in all
these investigations was limited to Gaussian or certain higher-order modes.
The low group-velocity of each transverse mode is different, resulting in the
dispersion of multi-mode profiles, and self-imaging may occur only at certain
distances \cite{ChengOL2007}. Waveguiding using an inhomogeneous medium was
studied for ultra-cold atoms in an anisotropic trap
\cite{VengalattorePRL2005,TarhanOL2007}. Nondiffracting spatial solitons of a
specific transverse shape may also be supported by self-focusing or
cross-focusing, due to a strong Kerr effect in EIT
\cite{HongPRL2003,FriedlerOL2005}.

Here, we analyze a novel scheme for spatial confinement in the paraxial
regime, which incorporates a large plane-wave pump, a uniform atomic spatial
distribution, and a weak probe, as opposed to the methods of finite pump,
finite atomic cloud, and Kerr solitons, respectively. Instead of imposing
transverse nonuniformity in real space, we prescribe nonuniformity in
$\mathbf{k}_{\perp}$ space, such that the paraxial optical diffraction, which
is also $\mathbf{k}_{\perp}$-dependent, is completely counterbalanced. Here,
$\mathbf{k}_{\perp}$ denotes the transverse wave-vectors, \textit{i.e.}, the
Fourier components of the envelope of the field in the transverse plane. We
study slow light via EIT in a dilute thermal vapor in the presence of a buffer
gas. Due to frequent velocity-changing collisions with the buffer gas atoms,
the atomic motion is diffusive, leading to the phenomena of Dicke narrowing
and diffusion of light \cite{FirstenbergPRA2008}. For a finite-sized probe and
a plane-wave pump, the atoms effectively "carry" the complex amplitude of the
probe field within their internal coherence as they diffuse, resulting in an
effective diffusion of the probe's envelope
\cite{PugatchPRL2007,ShukerImaging2007}. In this letter, we show that by
introducing a nonzero two-photon (Raman) detuning, the atomic motion also
induces a $\mathbf{k}_{\perp}$-dependence of the refraction index.
Specifically, for negative Raman detuning, the $\mathbf{k}_{\perp}$-dependent
refraction takes the shape of the paraxial diffraction with an opposite sign,
thus enabling its cancellation. This diffraction elimination is homogenous and
continuous, as opposed to discrete diffraction-management techniques
\cite{SilberbergPRL2000}.

\begin{figure}
[ptb]
\begin{center}
\includegraphics[
height=3.4048cm,
width=8.5141cm
]%
{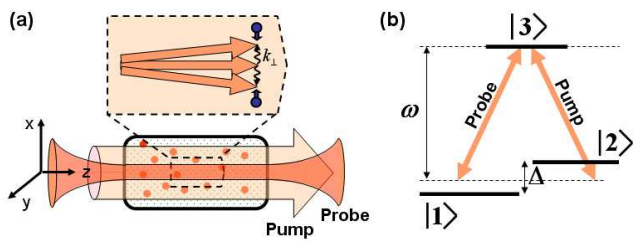}%
\caption{(a) Illustration of Doppler trapping of slow-light. A beamlike probe
and a plane-wave pump propagate along the $z$ direction. EIT effects in the
medium depend on the wave-vector difference $\mathbf{k}_{\perp},$ such that
atoms that move oppositely \ to $\mathbf{k}_{\perp}$ 'drag' the respective
field's component more efficiently back to the main axis. (b) Level structure.
To simplify the notation, the pump and the probe are assumed to have the same
frequency $\omega$, and the Raman detuning $\Delta$ is introduced via the
energy difference between the lower levels.}%
\label{fig_system}%
\end{center}
\end{figure}

The following simplified picture, illustrated in Fig. \ref{fig_system}(a),
explains this spatial-confinement phenomenon. Generally, for a negative
detuning, a moving atom couples more efficiently with the 'counter-propagating
components' (wave vectors) of the field due to the Doppler effect. In EIT, a
\emph{residual} Doppler effect takes place, which depends on the wave vector
associated with the difference between the pump and the probe
\cite{FirstenbergPRA2007,ShukerPRA2007}. In the simplest arrangement -- a
plane wave, degenerate, and co-propagating pump -- the pump-probe wave-vector
difference equals $\mathbf{k}_{\perp}$. Therefore, for negative Raman
detuning, each component in $\mathbf{k}_{\perp}$ space exhibits stronger
coupling with the atoms moving in the $(-\mathbf{k}_{\perp})$ direction and is
effectively carried back towards the main axis. This is, in fact, a
realization of a \emph{Doppler trapping of light by atoms}, in analogy with
the trapping of atoms in a Doppler optical trap.%

Consider a dilute vapor of $\Lambda$-type atoms, with two nearly-degenerate
lower states, $\left\vert 1\right\rangle $ and $\left\vert 2\right\rangle $,
and a single excited state $\left\vert 3\right\rangle $. A probe beam and a
strong plane-wave pump propagate along the $z$ direction, with equal frequency
$\omega,$ and couple states $\left\vert 1\right\rangle $ and $\left\vert
2\right\rangle $ with state $\left\vert 3\right\rangle $, respectively (see
Fig. \ref{fig_system}b). The energy difference between the lower levels,
$\Delta$, defines the Raman detuning \footnote{All the results are essentially
unchanged when a small pump-probe frequency difference is introduced.}, and
the one-photon detuning is assumed to be much smaller than the width of the
optical resonance. Under the paraxial approximation, assuming the changes in
the probe's envelope along the $z$ direction are much smaller than the changes
in the transverse plane, and assuming the pump is nearly constant along $z$,
the propagation of the probe in steady state can be described by
\cite{FirstenbergPRA2008},%
\begin{equation}
\left(  \frac{\partial}{\partial z}+i\frac{k_{\perp}^{2}}{2q}\right)
\Omega(\mathbf{k}_{\perp};z)=i\chi\left(  \Delta,\mathbf{k}_{\perp}\right)
\Omega(\mathbf{k}_{\perp};z), \label{prop1}%
\end{equation}
where $q=\omega/c$, $c$ is the speed of light, and $\chi\left(  \mathbf{k}%
_{\perp}\right)  $ is the linear susceptibility, with $\operatorname{Im}\chi$
being the absorption coefficient and $\operatorname{Re}\chi$ being dispersion.
$\Omega(\mathbf{k}_{\perp};z)$ is the Fourier transform of the slowly varying
Rabi envelope of the probe, defined by%
\begin{equation}
\Omega\left(  \mathbf{k}_{\perp};z\right)  =e^{i(\omega t-qz)}\int
d^{2}r_{\perp}e^{-i\mathbf{k}_{\perp}\cdot\mathbf{r}_{\perp}}\tilde{\Omega
}\left(  \mathbf{r}_{\perp};z,t\right)  , \label{prop2}%
\end{equation}
with $\tilde{\Omega}\left(  \mathbf{r}_{\perp};z,t\right)  $ the rapidly
oscillating Rabi frequency. The second summand in the left-hand side of
Eq.(\ref{prop1}) is the well-known diffraction term, which is quadratic in
$k_{\perp}$ and purely imaginary.

For an atom at rest, the susceptibility in the vicinity of the EIT line is
$\chi_{0}\left(  \Delta\right)  =i\alpha\lbrack1-\Gamma_{p}/(\Gamma
-i\Delta)],$ where $2\alpha$ is the absorption coefficient in absence of the
pump; $\Gamma_{p}$ is the power-broadening term, proportional to the pump
intensity; and $\Gamma$ is the total homogenous EIT line width -- the sum of
$\Gamma_{p}$ and the decoherence rate within the ground-state manifold. In a
vapor medium with buffer gas, the EIT atoms are subjected to frequent
velocity-changing collisions with the buffer-gas atoms \footnote{Spin-exchange
collisions between the EIT atoms are much less frequent and are taken into
account in the homogenous EIT width $\Gamma$.}. The resulting atomic motion is
diffusive and is characterized by a diffusion coefficient $D$, incorporating
both the mean thermal-velocity and the collision rate. Due to residual Doppler
broadening and Dicke narrowing, the EIT line shape becomes dependent on the
two-photon wave-vector difference, $\mathbf{k}_{\perp}\neq0$, and the
resulting susceptibility is \cite{FirstenbergPRA2008}
\begin{equation}
\chi\left(  \Delta,\mathbf{k}_{\perp}\right)  =i\alpha\left(  1-\frac
{\Gamma_{p}}{\Gamma+Dk_{\perp}^{2}-i\Delta}\right)  . \label{chi}%
\end{equation}
The term $Dk_{\perp}^{2}$ is the Doppler-Dicke width, originating from the
atomic motion \footnote{In Eq.(\ref{chi}), the parameters $\alpha,$
$\Gamma_{p},$ and $\Gamma$ are somewhat different from those for an atom at
rest and incorporate the atomic motion \cite{FirstenbergPRA2008}.}.

\begin{figure}
[ptb]
\begin{center}
\includegraphics[
height=6.3351cm,
width=8.4241cm
]%
{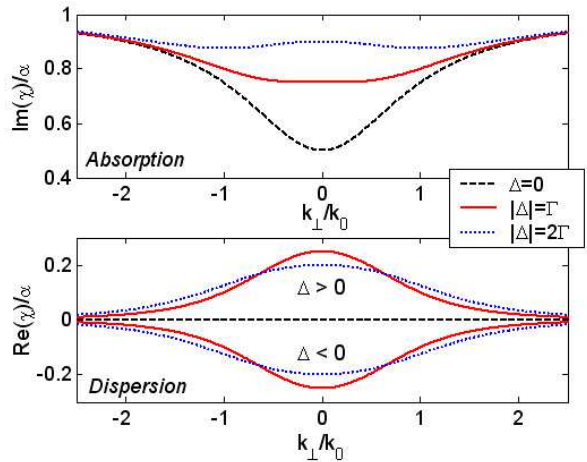}%
\caption{Imaginary (top) and real (bottom) components of the probe
susceptibility, normalized by the absorption in the absence of the pump,
$\alpha$, as a function of $k_{\perp}$ for different Raman detunings:
$\Delta=0$ (dashed-black), $\Delta=\pm\Gamma$ (solid-red), and $\Delta
=\pm2\Gamma$ (dotted-blue). The typical width is $k_{0}=(\Gamma/D)^{1/2},$
where $\Gamma=2\Gamma_{p}$ was chosen here. The negative quadratic shape of
the dispersion curve for $\Delta<0$, in the central $|k_{\perp}|$ region, can
be used to eliminate the optical diffraction.}%
\label{fig_chi}%
\end{center}
\end{figure}

On the Raman resonance, $\Delta=0$, the susceptibility $\chi$ is pure
imaginary and thus generates a $k_{\perp}$-dependent \emph{absorption filter}
without dispersion. The absorption filter for $\Delta=0$, depicted in Fig.
\ref{fig_chi} (dashed black), is a Lorentzian of width $k_{0}=(\Gamma
/D)^{1/2}.$ When the spatial spectra of the probe beam $\Omega\left(
\mathbf{k}_{\perp};z\right)  $ is confined within $k_{\perp}\ll k_{0},$ the
absorption is approximately quadratic in $k_{\perp}$ and, according to Eqs.
(\ref{prop1}) and (\ref{prop2}), operates as a Laplacian in real space, and
causes a diffusion-like behavior \cite{FirstenbergPRA2008}.%

For $k_{\perp}\ll k_{0}$, the absorption and the dispersion for nonzero Raman
detuning can be written as
\begin{subequations}
\begin{align}
\operatorname{Im}\chi &  =\operatorname{Im}\chi_{0}+\alpha\Gamma_{p}%
\Gamma\frac{\Gamma^{2}-\Delta^{2}}{\left(  \Gamma^{2}+\Delta^{2}\right)  ^{2}%
}\frac{k_{\perp}^{2}}{k_{0}^{2}}+O(k_{\perp}^{4}),\label{imx}\\
\operatorname{Re}\chi &  =\operatorname{Re}\chi_{0}-\alpha\Gamma_{p}%
\Gamma\frac{2\Gamma\Delta}{\left(  \Gamma^{2}+\Delta^{2}\right)  ^{2}}%
\frac{k_{\perp}^{2}}{k_{0}^{2}}+O(k_{\perp}^{4}). \label{rex}%
\end{align}
At the central part of the spatial spectrum, the dispersion is quadratic in
$k_{\perp},$ exactly like a paraxial diffraction term. Therefore, by properly
choosing the parameters, the dominant part of the motional-induced dispersion,
namely the $k_{\perp}^{2}$ term, can cancel the free-space diffraction. The
strength and the sign of the dispersion depend on $\Delta$, and, in order to
eliminate diffraction, a non-zero negative detuning is required. For the
specific case of $\Delta=-\Gamma,$ the absorption filter in Eq.(\ref{imx}) is
flat up to the fourth order in $k_{\perp}$, as seen in Fig. \ref{fig_chi}
(top, solid red), implying that no motional-induced diffusion will accompany
the propagation of paraxial images. By this we avoid spreading due to
absorption, which was significant, for example, in electromagnetically induced
focusing \cite{MoseleyPRA1996}. We therefore choose $\Delta=-\Gamma$ and,
following Eq. (\ref{prop1}), require the diffraction cancellation condition:
\end{subequations}
\begin{equation}
\frac{1}{2q}=\frac{\alpha\Gamma_{p}}{2\Gamma k_{0}^{2}}. \label{cond}%
\end{equation}

Under regular diffraction, a focused Gaussian beam that hits the medium with a
waist radius $w_{0}$ spreads as $w\left(  z\right)  ^{2}=w_{0}^{2}\left(
1+z^{2}/z_{R}^{2}\right)  $, with $z_{R}=qw_{0}^{2}/2$ being the Rayleigh
length. Condition (\ref{cond}) can be intuitively explained, by requiring the
diffraction spreading at one Rayleigh length ($w_{0}^{2}$) to be comparable to
the typical diffusion spreading ($D\tau$, where $\tau=z_{R}/v_{g}$ is the
slow-light delay and $v_{g}=\Gamma^{2}/(\alpha\Gamma_{p})$ is the group
velocity). For a waist of $\sim100$ $\mu$m and a Rayleigh length of a few cm,
conditions (\ref{cond}) and $k_{\perp}\ll k_{0}$ can be satisfied with $D$ of
the order of $10$ cm$^{2}$/s and $v_{g}$ of $\sim10$ km/s, which are readily
available \cite{PugatchPRL2007,ShukerImaging2007}. Notice however, that due to
the deviation from the Raman resonance-condition, the absorption per unit
length, $\kappa=2\alpha\lbrack1-\Gamma_{p}/(2\Gamma)]$, is substantial. It
becomes smaller as the power-broadening increases and eventually approach
$\kappa=\alpha$ for $\Gamma\approx\Gamma_{p}$. For a beam with $w_{0}%
=\pi/k_{0}$ and for $\Gamma\approx\Gamma_{p},$ condition (\ref{cond}) becomes
$\kappa=(\pi^{2}/2)/z_{R}\approx5/z_{R}$, which means the intensity decreases
by about $\exp\left(  -5\right)  $ every Rayleigh length.

In our scheme, strong absorption is unavoidable due to the non-zero Raman
detuning. While an observation of the non-diffraction phenomenon is well
within current experimental capabilities, applications of it may require
smaller absorption. Here, the fact that the absorption is independent of
$k_{\perp}$ is crucial, allowing a wide range of gain schemes to be
potentially applicable. Homogenous gain mechanisms that are available for
vapor, \textit{e.g.} Raman gain \footnote{See, \textit{e.g.}, J. A. Kleinfeld
and A. D. Streater, Phys. Rev. A \textbf{49}, R4301 (1994). By the addition of
another vapor species, with Raman gain properties, the absorption may be
compensated by controlling the relative atomic densities. The two-photon
bandwidth of the Raman resonance should be much larger than the Dicke
line-width of the image, to ensure the gain uniformity in $k_{\perp}$ space.},
can be considered, and specifically two integrated gain schemes in EIT were
recently explored \footnote{I. Novikova, D. F. Phillips, and R. L. Walsworth,
Phys. Rev. Lett. \textbf{99}, 173604 (2007). This gain is phase dependent.
However, we have numerically simulated its operation and showed that, thanks
to the non-diffraction process, the relevant phase is spatially conserved and
the gain is uniform.}\cite{WilsonGordonOL2008}. There is also the trivial
possibility to introduce gain before or after the cell, providing the gain
medium is much thinner than the Rayleigh length.%
\begin{figure}
[ptb]
\begin{center}
\includegraphics[
height=10.3241cm,
width=8.1429cm
]%
{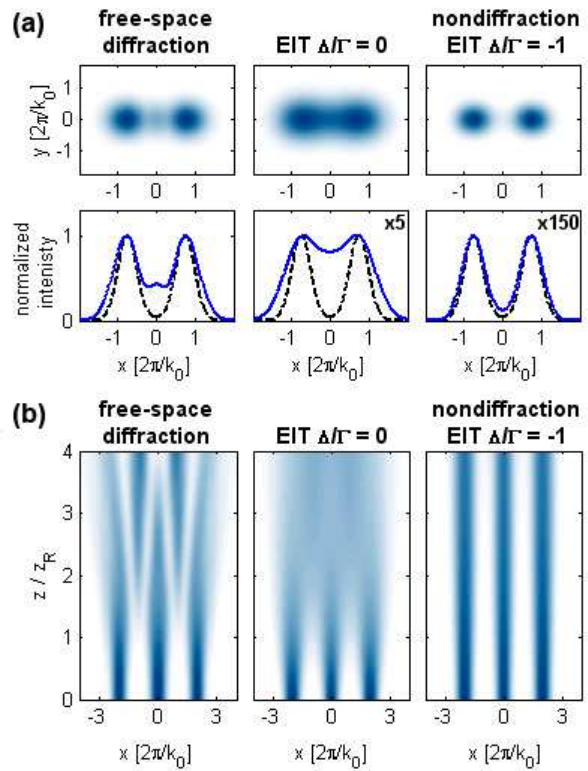}%
\caption{Numerical calculations demonstrating the effect of nondiffraction.
(a) An incident beam of two focused Gaussian modes, with a waist radius of
$w_{0}=\pi/k_{0}=100$ $\mu$m and separation of $3w_{0}$, propagating one
Rayleigh length for $\lambda=795$ nm. Condition (\ref{cond}) is satified,
$e.g.$, with $D=11$ cm$^{2}/$s and $v_{g}=9000$ m/s. The normalized
transmitted images and the profile cross-sections (incident is dashed,
transmitted is solid) are shown for three cases: free-space diffraction
(left); on-resonance EIT transmission (center); and EIT with a negative
detuning, $\Delta=-\Gamma$ (right), exhibiting no diffraction and no
diffusion. (b) Intensity at the $y=0$ plane (normalized for each $z$), of
three Gaussian beams with $4w_{0}$ separation, propagating $4$ Rayleigh
lengths.}%
\label{fig_effect}%
\end{center}
\end{figure}

\begin{figure}
[ptb]
\begin{center}
\includegraphics[
height=3.4553cm,
width=8.1319cm
]%
{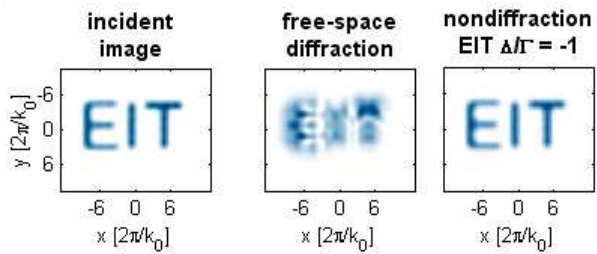}%
\caption{Numerical calculation of the nondiffraction effect for an image of
the letters 'EIT' that traverses twice the typical Rayleigh length ($\pi
/k_{0}=50$ $\mu$m, $\lambda=795$ nm).}%
\label{fig_images}%
\end{center}
\end{figure}

Figure \ref{fig_effect} presents numerical calculations of the effect,
obtained by taking the Fourier transform of the boundary condition,
$\Omega\left(  x,y;z=0\right)  $, according to Eq. (\ref{prop2}), solving Eq.
(\ref{prop1}), and doing the inverse Fourier transform. The exact expression
(\ref{chi}), rather than the approximation of Eqs. (\ref{imx}) and
(\ref{rex}), was used for the calculation. Figure \ref{fig_effect}(a)
demonstrates the transmission of two Gaussian beams in the transverse plane at
$z_{R}$, and Fig. \ref{fig_effect}(b) depicts the propagation of three beams
along the $z$ axis up to $4z_{R}$. Without EIT (left column), there is only
free-space diffraction, and with EIT on-resonance ($\Delta/\Gamma=0,$ center
column), the diffraction spreading is accompanied by a diffusion spreading,
due to the $k_{\perp}$-dependence of $\operatorname{Im}\chi$. For EIT with
$\Delta/\Gamma=-1$ (right column), the elimination of spreading due to both
diffraction and diffusion is clearly evident. Notice that the width of each
Gaussian beam in Fig. \ref{fig_effect}(b) increases by $\sim50\%$ after
$4z_{R}$. This is the effect of fourth order in $\chi\left(  k_{\perp}\right)
,$ and we have verified numerically that the spreading after $4$ Rayleigh
lengths approaches zero as $w_{0}$ is increased (\textit{e.g.}, for a waist of
$w_{0}=8\times\pi/k_{0},$ the spreading is $\sim3\%$). An example of
nondiffraction of an elaborated image that traverses $2z_{R}$ is presented in
Fig. \ref{fig_images}. As evident from Figs. \ref{fig_effect} and
\ref{fig_images} and in contrast to previous nondiffraction schemes, our
scheme works for a \emph{general} image and for \emph{any distance} along the
propagation direction.%

In conclusion, we utilize the EIT linear susceptibility in the wave-vector
space, rather than in real space, to eliminate the diffraction of a paraxial
probe beam with a general transverse profile, limited in $k_{\perp}$-space to
the region $k_{\perp}\ll(\Gamma/D)^{1/2}$. From the viewpoint of optical
information processing, our scheme may be useful to increase the capacity of
information carried by the slow-light and hence also the memory capacity in
storage of light. As $\Gamma$ is increased and $D$ is decreases, the
resolution of the nondiffracting pattern may be increased. Elongated narrow
beams can also be utilized for the purpose of guiding, for example, via
non-linear interactions or dipole trapping. An intriguing extension of this
work would be to generalize the two-dimensional 'Doppler trap' to pulses of
finite duration, in order to achieve trapping in three dimensions.

\bibliographystyle{apsrev}

\end{document}